\documentclass[aps,pra,floatfix,notitlepage,twocolumn,preprintnumbers,showpacs,longbibliography,nofootinbi,superscriptaddress]{revtex4-1}

\usepackage{inputenc}
\usepackage{natbib}
\usepackage[colorlinks=true,citecolor=NavyBlue,urlcolor=NavyBlue,linkcolor=NavyBlue]{hyperref}

\usepackage{graphicx}
\usepackage{float}
\usepackage[english]{babel}
\usepackage{amsmath}
\usepackage[normalem]{ulem}
\usepackage{bbold}
\usepackage{physics} 
\usepackage[dvipsnames]{xcolor}
\usepackage{enumitem}
\usepackage{amssymb}

\allowdisplaybreaks

\begin{document}

\title{
Atom-Photon Bound States in Fractal Photonic Lattices:\\Localization Length and Anomalous Diffusion
}

\author{Florian B\"onsel}
\affiliation{Max Planck Institute for the Science of Light, 91058 Erlangen, Germany}
\affiliation{Department of Physics, Friedrich-Alexander-Universit\"at Erlangen-N\"urnberg, 91058 Erlangen, Germany}

\author{Flore K.~Kunst}
\affiliation{Max Planck Institute for the Science of Light, 91058 Erlangen, Germany}
\affiliation{Department of Physics, Friedrich-Alexander-Universit\"at Erlangen-N\"urnberg, 91058 Erlangen, Germany}

\author{Federico Roccati}
\affiliation{Max Planck Institute for the Science of Light, 91058 Erlangen, Germany}
\affiliation{Quantum Theory Group, Dipartimento di Fisica e Chimica -- Emilio Segr\`e,
    Universit\`a degli Studi di Palermo, via Archirafi 36, I-90123 Palermo, Italy}

\begin{abstract}
We study atom-photon bound states seeded by two-level emitters coupled to self-similar photonic lattices. By expressing the photonic Green's function through the heat kernel, we show that the far-field localization length obeys $\xi \sim \Delta^{-1/d_w}$, with the detuning $\Delta$ from the lower spectral edge and the walk dimension $d_w$ of the underlying fractal. This scaling is controlled by anomalous diffusion and does not rely on translational invariance or a band-edge effective-mass approximation. Exact diagonalization on Sierpi\'nski gaskets, pyramids, Vicsek graphs, and Sierpi\'nski carpets confirms the far-field prediction once the bath Hamiltonian is rendered Laplacian-like by compensating the local inhomogeneity in the connectivities with on-site potentials. In the near field, the bound-state amplitude exhibits an additional algebraic variation. For nested finitely ramified fractals, the corresponding exponent agrees with the classical resistance/ first-passage scaling, whereas Sierpi\'nski carpets display clear deviations from this simple law. Our results extend structured-bath waveguide QED to self-similar non-periodic geometries and connect bound-state profiles to transport exponents of the underlying fractal lattice.
\end{abstract}
\maketitle

\begin{figure*}[htb!]
		\centering
\includegraphics[width=\textwidth]{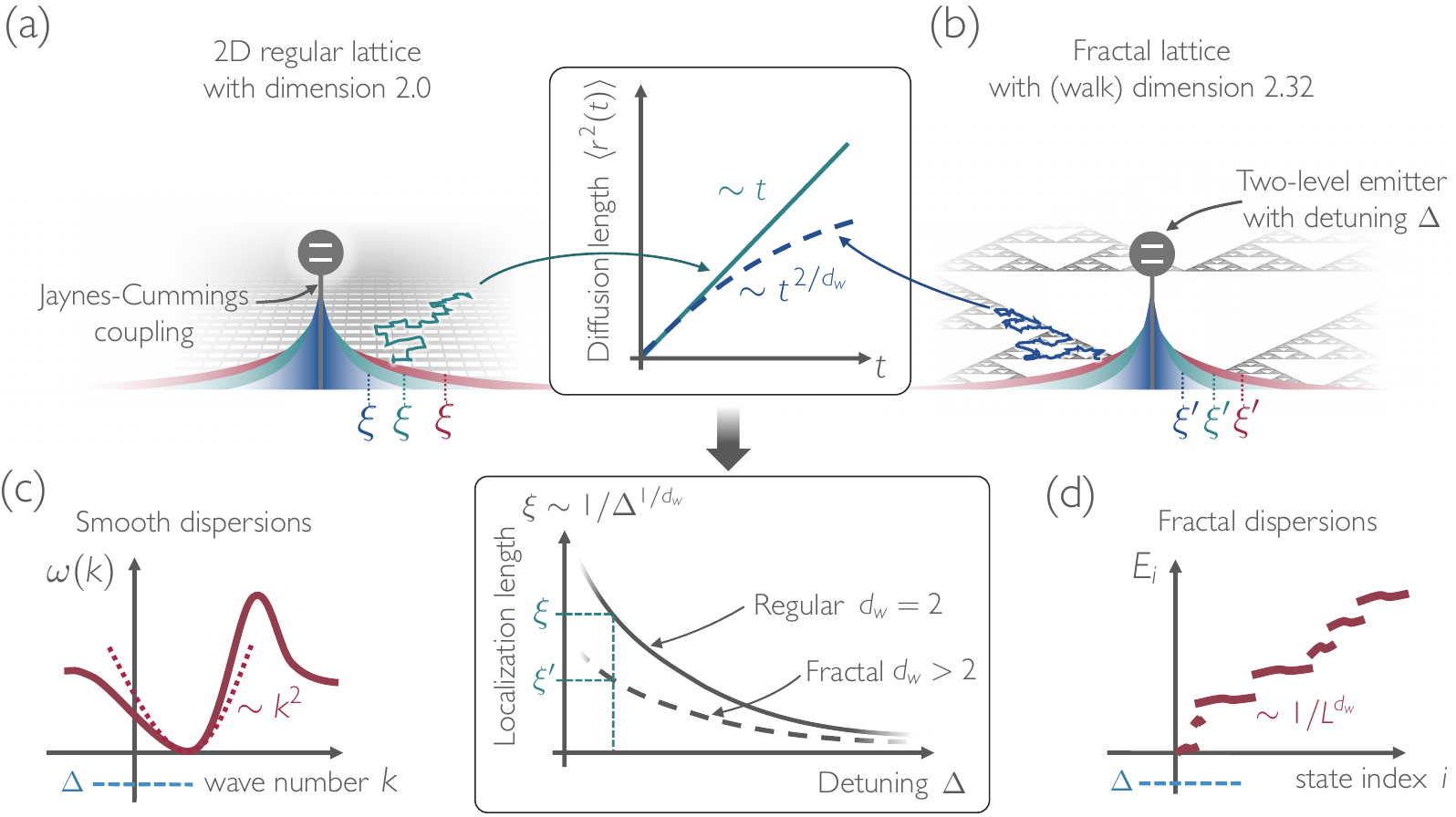}
		\caption{\textit{Atom-photon bound states}. Quantum emitters coupled to a regular (a) and fractal (b) photonic lattice. (c) In the regular case, a quadratic band edge $\omega(\mathbf{k})-\omega_{\rm edge}\sim |\mathbf{k}|^2$ yields the familiar scaling $\xi\sim \Delta^{-1/2}$ for the localization length $\xi$ of the atom-photon bound state, with $\Delta$ as the emitter detuning from the lower band edge. (d) In a fractal bath, translational invariance and ordinary bands are absent; instead the low-energy scaling is set by the walk dimension through $E\sim L^{-d_w}$, leading to the generalized law $\xi\sim \Delta^{-1/d_w}$, with the walk dimension $d_w$ (defined in the text). For a regular two-dimensional lattice, $d_w=2$ (and coincides with the spatial dimension), whereas for the Sierpi\'nski gasket, $d_w\approx 2.32$.}
		\label{fig:illustration}
	\end{figure*}

%\section{Introduction}\label{sec:one}
Waveguide quantum electrodynamics (QED) has emerged as a powerful platform for controlling light-matter interactions at the quantum level, driven by the ability to engineer structured photonic baths, such as coupled-resonator arrays, photonic crystal waveguides, and circuit-QED lattices, providing desired bandgaps and dispersion relations~\cite{Goban2014, Lodahl2015, Douglas2015, VanLoo2013, sheremet2023waveguide, ciccarello2024waveguide}. A typical feature of these systems is the formation of atom-photon bound states, which can be understood as a quantum impurity problem in the photonic bath: hybrid excitations with the photonic component remaining localized around a quantum emitter whose transition frequency lies inside a photonic bandgap~\cite{john1990quantum, liu2017quantum, Calajo2016, gonzalez2017quantum, verstraten2025control, leonforte2021dressed, shi2016bound}. 
These bound states can mediate coherent, long-range emitter-emitter interactions and allow a wide range of protocols for entanglement distribution, effective spin Hamiltonians, and the simulation of quantum many-body effects~\cite{Mirhosseini2020, Sundaresan2019, AsenjoGarcia2017, roccati2025many, hansen2026realization}.

In translationally invariant lattices, the properties of bound states are well understood~\cite{john1991quantum, raman2010photonic, john1994spontaneous}. Near a band edge, the photonic dispersion is parabolic, yielding an effective mass and a localization length $\xi$ that diverges as $\xi \sim \Delta^{-1/2}$ with $\Delta$ being the atom-bath detuning from the band edge. Fractal photonic lattices, such as the famous Sierpi\'nski fractals, lack translational invariance, making standard concepts like momentum space or energy bands ill-defined~\cite{alexander1982density, barlow2006diffusions, Telcs2006}.
These self-similar structures exhibit non-integer Hausdorff dimension and support anomalous diffusion characterized by the spectral dimension $d_s < 2$ and walk dimension $d_w > 2$ (both defined later). Classical random walks spread sub-diffusively, as measured by the mean squared displacement $\langle r^2(t)\rangle \sim t^{2/d_w}$ with a return probability $\sim t^{-d_s/2}$. Such power-law responses and fractal-like memory effects are also central to describing the dynamics of active particles in viscoelastic media, with fractional calculus providing a natural framework for anomalous diffusion behavior~\cite{quevedo2025active}.

Over the past decade, fractal lattices have been realized experimentally via femtosecond-laser-written waveguide arrays and explored theoretically, revealing flat bands and compact localized states~\cite{Mukherjee2018, hanafi2021localized}, anomalous Floquet topological phases~\cite{li2023fractal, yang2020photonic}, higher-order topological corner modes~\cite{Ren2023}, and anomalous quantum transport~\cite{xu2021quantum, darazs2014transport, rojo2024anomalous}.

Departing from strict periodicity opens new frontiers for controlling light-matter interactions. Extensive research into photonic quasicrystals and complex aperiodic structures has revealed rich optical phenomena utilized for novel material properties \cite{vardeny2013optics, boriskina2015making}. Within periodic but topologically non-trivial or structured baths, atom-photon bound states with unusual chiral and long-range properties have been predicted~\cite{bello2019unconventional,
leonforte2024quantum, Bonsel2026fibonacciwaveguide}. 

Despite these advances in fractal and aperiodic photonics, the impact of such baths on quantum impurity problems and atom-photon bound states in particular, remains largely unexplored. An early step in this  direction considered spontaneous emission of a quantum emitter coupled to a QED vacuum with a fractal energy spectrum, finding power-law rather than exponential decay~\cite{akkermans2013spontaneous}. The general question of how a spatially fractal photonic bath reshapes the  atom-photon bound state -- its localization length and spatial profile -- has not been addressed. Given the absence of standard tools like the Bloch theorem for non-periodic environments, we tackle the following question: \textit{How does a fractal photonic bath reshape atom-photon bound states?}

In this work, we show that the central and most robust consequence of a fractal bath is a localization length $\xi$ that diverges as $\xi \sim \Delta^{-1/d_w}$ when the emitter approaches the spectral edge. The walk dimension $d_w$, that is the exponent controlling anomalous diffusion on the graph, therefore replaces the band curvature of a translationally invariant lattice. The spectral dimension $d_s$ enters instead through a subleading algebraic structure of the bound state close to the emitter. For the nested finitely ramified fractals studied here, this near-field exponent reduces to the classical resistance/ first-passage exponent; for infinitely ramified carpets we use that prediction as a benchmark and find systematic deviations.

We derive the scaling analytically by expressing the photonic Green's function at the bound-state energy through its Laplace transform in terms of the classical heat kernel~\cite{economou2006green, barlow2006diffusions}. Established sub-Gaussian heat kernel bounds on fractals~\cite{BarlowPerkins1988, barlow2005characterization, kumagai2014random, barlow2010uniqueness} then enable a controlled saddle-point analysis of the far-field regime. Numerically, we validate the prediction on canonical examples when the underlying fractal graphs are made Laplacian-like. This is done via adding degree-dependent on-site energies with the site degree $d_i$ defining the number of connections of site $i$.

These results position fractal photonic lattices as a distinct waveguide-QED setting in which transport exponents and ramification properties replace band curvature as the relevant control parameters.
Throughout this work, the terms ``near-field'' and ``far-field'' do not refer to the usual radiative zones of waveguide QED. Instead, they denote two distance regimes of the atom-photon bound-state profile measured via the \textit{chemical}  distance  (defined later) from the emitter. The far-field regime is the asymptotic tail, $r\gg \xi$, at which the wavefunction is exponentially suppressed and the localization length $\xi$ can be extracted. The near-field regime is the short-distance regime, $r\ll \xi$, more precisely $r^{d_w}\Delta\ll 1$, at which the exponential factor is not yet operative and the spatial dependence is governed by the underlying fractal diffusion.

We organize the paper as follows. In Section~\hyperref[sec:two]{I}, we introduce the basic model, the notation, and the standard description of atom-photon bound states in regular lattices. In Section~\hyperref[sec:three]{II}, we summarize the self-similar fractal lattices considered here. In Section~\hyperref[sec:four]{III}, we reformulate the problem in terms of the heat kernel and derive the far-field and near-field scaling laws. In Section~\hyperref[sec:five]{IV}, we present the numerical results for nested finitely and infinitely ramified fractals. Lastly, we draw our conclusions in Section~\hyperref[sec:six]{V}.

\section{Atom-photon bound states -- Standard formalism}\label{sec:two}
We consider a single two-level system (TLS) with ground state \(|g\rangle\) and excited state \(|e\rangle\) and transition frequency \(\omega_e\) coupled to a $D$-dimensional ($D$ integer) translationally invariant photonic bath. 

The full system Hamiltonian in the rotating-wave approximation ($g\ll |\Delta|,\omega_e$) reads
\begin{equation}\label{eq:H}
    \hat H 
    = 
    \omega_e \hat\sigma^+\hat\sigma^- 
    + 
    \sum_{\mathbf{k}} 
    \omega_{\mathbf{k}} 
    \hat a_{\mathbf{k}}^\dagger \hat a_{\mathbf{k}}
    + 
    \sum_{\mathbf{k}} 
    \left( g_{\mathbf{k}} \hat\sigma^+ \hat a_{\mathbf{k}} + \text{H.c.} \right),
\end{equation}
with \(\hat\sigma^+=|e\rangle\langle g|\) and \(\hat\sigma^- = |g\rangle\langle e|\) the atomic transition operators, $\omega_{\mathbf{k}}$ the bath dispersion, and $g_{\mathbf{k}}$ the coupling between the TLS and bath mode $\mathbf{k}$. We write the emitter frequency as $\omega_e=\omega_{\rm edge}-\Delta$ and focus on bound states with energy $E_{\rm BS}$ lying in the gap below the lower bath band edge.

In the single-excitation sector, which is our focus in the entire manuscript, we can apply the general ansatz for an eigenstate with energy \(E_{\rm BS}\)
\begin{equation}\label{eq:ansatz}
|\Psi_{E_{\rm BS}}\rangle = \left( c_e\,\sigma^+ + \sum_{\mathbf{k}} \phi_{\mathbf{k}}\,a_{\mathbf{k}}^\dagger \right)|g;{\rm vac}\rangle,
\end{equation}
and the normalization constraint \( |c_e|^2 + \sum_{\mathbf{k}} |\phi_{\mathbf{k}}|^2 =1\). Inserting \eqref{eq:ansatz} into the time-independent Schr\"odinger equation \(\hat H|\Psi_{E_{\rm BS}}\rangle = E_{\rm BS} |\Psi_{E_{\rm BS}}\rangle\) yields 
the pole equation for the bound state energy~\cite{Calajo2016}: $E_{\rm BS}-\omega_e - \Sigma(E_{\rm BS}) = 0$ with $ \Sigma(z) \equiv \sum_{\mathbf{k}} |g_{\mathbf{k}}|^2/ (z-\omega_{\mathbf{k}})$ as the emitter's self-energy. The real-space photonic wavefunction amplitude of the bound state at position $\mathbf{r}$ is given by the Fourier transform
\begin{equation}\label{eq:psi-r-G-general}
\psi(\mathbf{r}) = \sum_{\mathbf{k}} \phi_{\mathbf{k}} e^{i \mathbf{k}\cdot \mathbf{r}}
= c_e \sum_{\mathbf{k}} \frac{g_{\mathbf{k}}^* e^{i \mathbf{k}\cdot \mathbf{r}}}{E_{\rm BS} - \omega_{\mathbf{k}}}.
\end{equation}
Near the band edge, the variation of $g_{\mathbf{k}}$ with $\mathbf{k}$ is usually small, allowing us to write $g_{\mathbf{k}} \approx g$.
Therefore, writing the bath's Green's function
\begin{equation}\label{eq:Greens_functin_plane_waves}
G(\mathbf{r};E) \equiv \sum_{\mathbf{k}} \frac{e^{i \mathbf{k}\cdot \mathbf{r}}}{E - \omega_{\mathbf{k}}},
\end{equation}
the final expression of the bound state reads
$\psi(\mathbf{r}) = g\, c_e\, G(\mathbf{r};E_{\rm BS})$
with energy satisfying $E_{\rm BS} - \omega_e - g^2 G(0;E_{\rm BS}) = 0$.
For weak coupling, we have 
\begin{equation}\label{eq:E_BS_E_edge_minus_Delta}
E_{\rm BS} \approx \omega_e = \omega_{\rm edge}-\Delta.
\end{equation}

%\subsubsection{Band-edge approximation}
For an ordinary band edge in a regular lattice, the dispersion can be expanded to quadratic order as
$\omega_{\mathbf{k}} \simeq \omega_{\rm edge} + a |\mathbf{k}|^2$, with $a>0$.
Using Eq.~\eqref{eq:E_BS_E_edge_minus_Delta}, we define
$\kappa \equiv (\Delta/a)^{1/2} > 0$,
so that
$E_{\rm BS} - \omega_{\mathbf{k}} \approx -a\left(|\mathbf{k}|^2 + \kappa^2\right)$.
In the continuum limit,  and after angular integration, the Green's function becomes
\begin{equation}\label{eq:G-general_real_space}
G(\mathbf{r};E_{\rm BS}) \propto \int_0^\infty dk \, \frac{k^{D-1} J_{\frac{D}{2}-1}(k r)}{(k r)^{D/2-1} (k^2 + \kappa^2)},
\end{equation}
with $r=|\textbf{r}|$ and $k=|\textbf{k}|$, whose closed form is  $G(r) \propto \left( \kappa/r \right)^{D/2-1} K_{D/2-1}(\kappa r)$~\cite{economou2006green}
with  $J_\nu$ the Bessel function of the first kind and $K_\nu$ the modified Bessel function of the second kind. Therefore, at large distance \(r\gg \xi\),
\begin{equation}
\psi(r) \propto g\, c_e\, \frac{e^{-\kappa r}}{r^{(D-1)/2}},
\end{equation}
and the localization length obeys
\begin{equation}\label{eq:xi_band_edge}
\xi = \kappa^{-1} \sim \Delta^{-1/2}.
\end{equation}
This standard result will serve as the reference point for the fractal generalization developed below.

\section{Self-similar photonic fractals}\label{sec:three}
In this work, each photonic lattice is 
described by the quadratic bosonic Hamiltonian
\begin{equation}
    \hat H_\text{bath} 
    =
    -J
    \sum_{i,j}
    A_{ij}
    \hat a_{\mathbf{r}_i}^\dagger \hat a_{\mathbf{r}_j}
\end{equation}
with $A=\{ A_{ij}\}$ the adjacency matrix of  a finite graph 
$G=(V,E)$, with vertices $V$ corresponding to resonators and edges $E$ encoding nearest-neighbor hopping. All physical observables are defined on the graph, and distances between two sites $\mathbf r$ and $\mathbf r'$ are measured by the chemical, or shortest-path, distance $d(\mathbf r,\mathbf r')$.
For example, if two sites at $\mathbf{r}$ and $\mathbf{r}'$ have a shortest path of three connections, then $d(\mathbf{r},\mathbf{r'})=3$. For sites along straight lines on the graph which are mutually connected, one gets $d(\mathbf{r},\mathbf{r'})=|\mathbf{r'}-\mathbf{r}|$.

The fractal lattices considered here are generated iteratively from a finite building block and by applying a substitution rule. Each iteration defines a generation $g$, which controls the system size and the number of lattice sites. Increasing the generation produces larger graphs that approximate the infinite fractal in the thermodynamic limit.

Throughout this work, we consider two classes of self-similar lattices~\cite{hutchinson1981fractals, falconer1985geometry}: nested finitely ramified fractals and infinitely ramified fractals. While both exhibit exact geometric self-similarity, their connectivity properties differ in a fundamental way with important consequences for transport and localization.
\subsection{Nested fractals}
Nested fractals are finitely ramified graphs, meaning that the removal of a finite number of vertices is sufficient to disconnect the lattice at any scale. This is known as the nesting theorem~\cite{lindstrom1990brownian}. Canonical examples include the Sierpi\'nski gasket, Vicsek graph, and higher-dimensional generalizations of the gaskets. These lattices possess a recursive structure that allows for an exact decimation into smaller self-similar copies that intersect only at finitely many points, see Fig.~\ref{fig:fractal_generations}.

Fig.~\ref{fig:fractal_generations} illustrates representative generations of the nested finitely ramified and infinitely ramified lattices studied in this work. For the Sierpi\'nski gaskets, we introduce the parameter $b$. A value of $b=2$ means that the generation $g$ of the fractal is created by placing the $g-1$ generation at the corners of a triangle with an edge made from $b=2$ sites, as in Fig.~\ref{fig:fractal_generations}(a), whereas a $b=3$ Sierpi\'nski gasket is shown in Fig.~\ref{fig:fractals}. The Sierpi\'nski carpets are characterized by two numbers $(m,n)$ meaning that the central void is made by cutting $n$ pieces out of the total $m$ pieces of the fractal. While Fig.~\ref{fig:fractal_generations}(b) shows the $(9,1)$ Sierpi\'nski carpet, the $(16,4)$ carpet is shown in Fig.~\ref{fig:fractals}. For nested fractals, the connectivity pattern remains self-similar under rescaling. As a result, these structures admit exact renormalization schemes~\cite{hilfer1984renormalisation} and have well-characterized geometric and transport exponents~\cite{BarlowPerkins1988}.

\subsection{Infinitely ramified fractals}
In contrast, infinitely ramified fractals cannot be disconnected by removing any finite number of vertices. The paradigmatic example is the Sierpi\'nski carpet, constructed by recursively removing subsets of sites from a regular lattice. Carpets contain voids at all length scales and exhibit a highly heterogeneous connectivity.

\begin{figure}
		\centering
\includegraphics[width=\columnwidth]{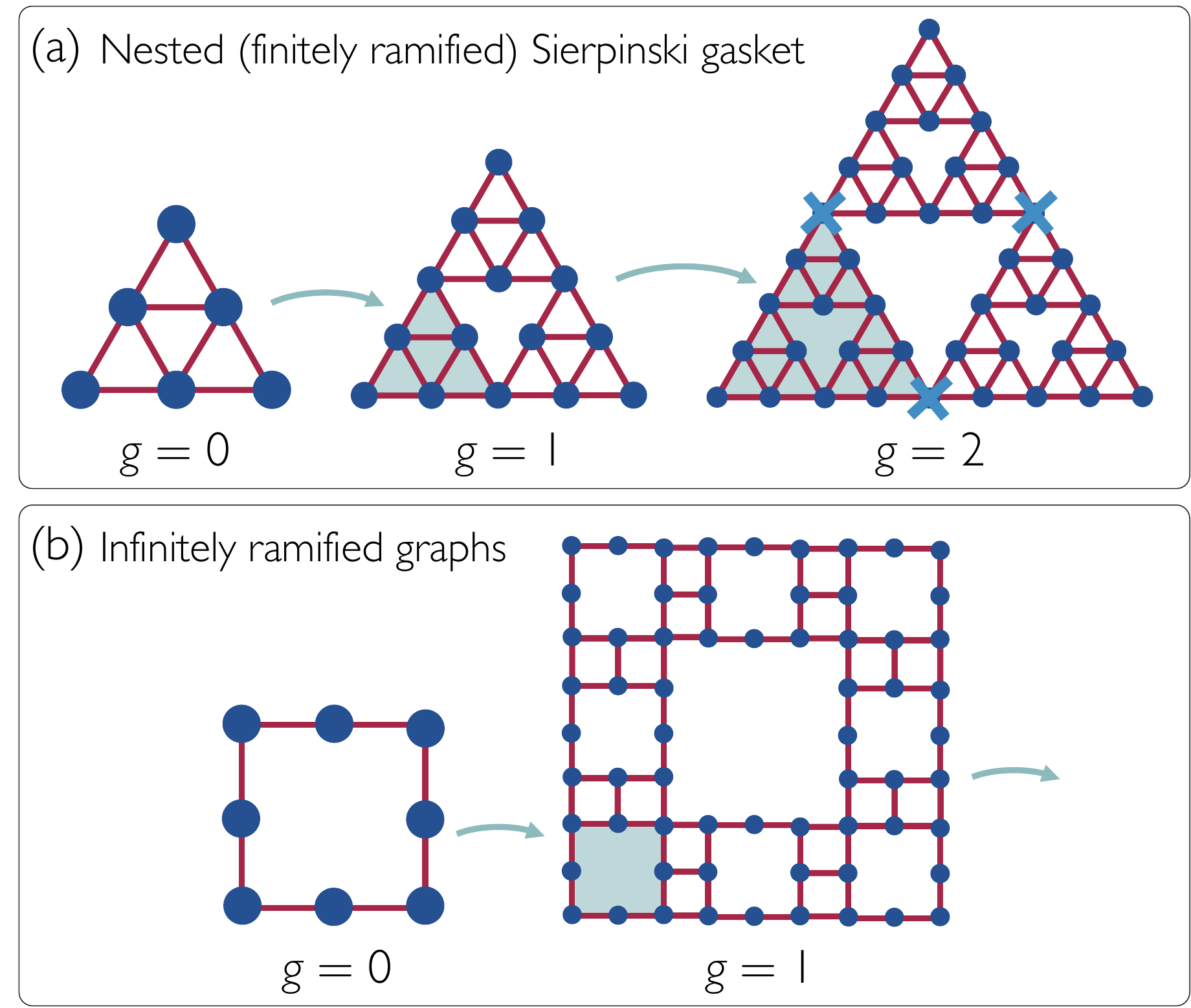}
		\caption{\textit{Representative self-similar photonic graphs}. Successive generations $g$ illustrate how nested finitely ramified (a) and infinitely ramified (b) fractals are created by iterating a local substitution rule. Nested fractals, such as Sierpi\'nski gaskets, can be disconnected by removing finitely many vertices at each scale, here three sites for the $b=2$ gasket [blue crosses for $g=2$ in (a)], whereas Sierpi\'nski carpets remain connected across arbitrarily many channels.}
		\label{fig:fractal_generations}
	\end{figure}

\begin{figure*}
		\centering
\includegraphics[width=\textwidth]{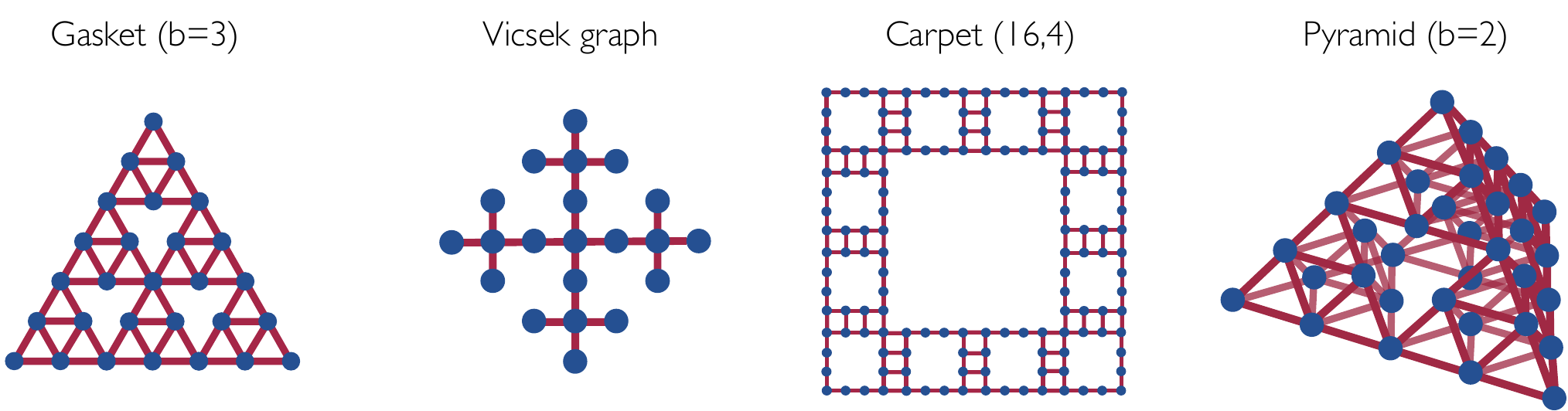}
		\caption{\textit{Fractal photonic lattices}. Graphs of second generation ($g=2$) fractals considered in this work, along with the ones displayed in Fig.~\ref{fig:fractal_generations}.}
		\label{fig:fractals}
	\end{figure*}

Fig.~\ref{fig:fractal_generations} also shows representative Sierpi\'nski carpets. While geometric self-similarity is preserved, the connectivity structure is qualitatively different from that of nested finitely ramified lattices because transport can proceed through arbitrarily many channels at every scale.

\section{Atom-photon bound states in photonic fractals}\label{sec:four}
For fractals it is well established that diffusion is anomalous and that excitations typically scale with a characteristic length scale $L$ as $E \sim L^{-d_w}$~\cite{argyrakis1992density, nakayama1994dynamical, alexander1982density, rammal1983random, gefen1983anomalous, o1985diffusion, o1985analytical}. Identifying the localization length $\xi$ as the characteristic length scale of a bound state, this dynamical scaling relation yields $\Delta \sim \xi^{-d_w}$. Consequently, this argument suggests a generalized version of \eqref{eq:xi_band_edge} as
\begin{equation}
    \xi\sim \Delta^{-1/d_w}.
\end{equation}
Exact results on localization in fractal lattices support this 
picture~\cite{wang1995localization, li2026anderson}. However, this scaling with the non-integer parameter $d_w$ is a specific property of the Laplacian matrix of an underlying graph. A generic tight-binding Hamiltonian on a fractal does not necessarily coincide with the Laplacian due to fluctuations in the local coordination number, i.e., the number of nearest neighbors of each site. Consequently, one cannot naively assume that $E(k) \sim k^{d_w}$ holds for an arbitrary lattice, nor is the momentum-space integral well-defined since $k$ is not a conserved quantum number. To derive the bound state properties, we formulate the problem entirely in real space.

Defining the shifted bath Hamiltonian matrix $\tilde{H}_\text{bath} = H_\text{bath}  - E_{\min} \ge 0$, with $E_\text{min}$ the lowest eigenvalue of $H_\text{bath}$, and the positive detuning $\Delta = E_{\min} - E_{\rm BS}$, we rewrite the Green's function as
$G(\mathbf{r}) = \langle \mathbf{r}| (E_{\rm BS} - H)^{-1} | \mathbf{0} \rangle
    = - \langle \mathbf{r}| (\tilde{H} + \Delta)^{-1} | \mathbf{0} \rangle$. 
Using the Laplace-transform identity $\hat{A}^{-1} = \int_0^\infty d\tau\, e^{-\hat{A}\tau}$ for a positive definite operator $\hat{A}$, we obtain
\begin{equation}\label{eq:resolvent_heat_kernel}
    G(\mathbf{r}) = -\int_0^\infty d\tau\, e^{-\Delta\tau} K(\mathbf{r}, \tau).
\end{equation}
Here, we identified the heat kernel $K(\mathbf{r},\tau) \equiv \langle \mathbf{r}|e^{-\tilde{H}\tau}|\mathbf{0}\rangle$, which describes diffusion on the underlying graph.

The heat kernel estimates invoked below apply to the graph Laplacian $L=D-A$ with $D$ the diagonal degree matrix telling how many connections each site has and $A$ the adjacency matrix. By contrast, a standard photonic tight-binding Hamiltonian matrix usually takes the form $H_{\mathrm{bath}}=-J A$,  with $J$ being the hopping rate. To render the Hamiltonian Laplacian-like, we need to add on-site energies to each site $i$ as $d_iJ$. For example, the Sierpi\'nski gasket ($b=3$) has sites with degrees $2$, $4$ and $6$, whereas the Vicsek fractal has sites with degree $1$, $2$, and $4$, cf. Fig.~\ref{fig:fractals}. In other words, the fractal baths are self-similar, but locally inhomogeneous in the site-degree numbers. We compensate this inhomogeneity with additional on-site terms to extract the geometric impact  of the fractal structure on the localization properties. This step is central for our comparison between analytics and numerics. With the Hamiltonian properly rendered Laplacian, the heat kernel $K(\mathbf{r}, \tau)$ for a wide class of nested fractals satisfies established sub-Gaussian bounds valid for nested finitely ramified fractals, such as Sierpi\'nski gaskets or Vicsek graphs ~\cite{BarlowPerkins1988, Kumagai1993, hambly1999transition, 
barlow2005characterization, kumagai2014random}, and for infinitely ramified structures, including Sierpi\'nski carpets \cite{barlow2010uniqueness}. Specifically, for times $\tau > d(\mathbf{r})$, the kernel is bounded by
\begin{equation}\label{eq:sub_gaussian_kernel_bounds}
    K_l(\mathbf{r}, \tau)
    \leq 
    K(\mathbf{r}, \tau) 
    \leq 
    K_r(\mathbf{r}, \tau)
\end{equation}
with
\begin{equation}
    K_\alpha(\mathbf{r}, \tau)
    =
    \frac{c_\alpha}{\tau^{d_s/2}}  \exp \left[ - C_{\alpha} \left( \frac{d(\mathbf{r})^{d_w}}{\tau} \right)^{1/(d_w-1)} \right]
    , 
\end{equation}
$\alpha = l,r$, and $d(\mathbf{r})$ is the chemical distance from the origin, which will be the site at which the emitter couples to, and $C_{r,l}, c_{r,l}$ are positive numerical constants. These bounds highlight that the diffusion dynamics are governed by two exponents distinct from the fractal dimension $d_f$:
(i) the \textit{walk dimension} \(d_w\). This characterizes how fast a random walker spreads on a structure through the anomalous diffusion law
\(\langle r^2(t)\rangle \sim t^{2/d_w}\) \cite{havlin1987diffusion, ben2000diffusion}.
On Euclidean lattices, \(d_w = 2\), yielding standard Brownian motion \(\langle r^2\rangle \sim t\), whereas on fractals typically \(d_w > 2\), indicating subdiffusive transport;
(ii) the \textit{spectral dimension} \(d_s\). This governs the low-energy density of states through the scaling \(\rho(E) \sim E^{d_s/2 - 1}\) as \(E \to 0\) \cite{alexander1982density}, as well as the return probability of a random walker \(P_{\mathrm{return}}(t) \sim t^{-d_s/2}\) \cite{rammal1983random}. Generally, \(d_s < d_f\), reflecting that vibrational and diffusive modes effectively explore a space of lower fractal dimension than the geometric embedding.

\subsection{Far-field scaling}
The resolvent integral~\eqref{eq:resolvent_heat_kernel} is bounded between two integrals by the sub-Gaussian bounds~\eqref{eq:sub_gaussian_kernel_bounds}. 
We write
\begin{equation}\label{eq:int_1}
  \begin{aligned}
       -c_r\int_0^\infty d\tau\, \tau^{-d_s/2}\, e^{-S_r(d(\mathbf{r}),\tau)} 
   \le  G(d(\mathbf{r})) \\
   \le -c_l\int_0^\infty d\tau\, \tau^{-d_s/2}\, e^{-S_l(d(\mathbf{r}),\tau)},
  \end{aligned}
\end{equation}
and introduced the action
\begin{equation}
    S_{l,r}(d(\mathbf{r}),\tau) = \Delta\tau + C_{l,r}\,\left(\frac{d(\mathbf{r})^{d_w}}{\tau}\right)^{1/(d_w-1)}.
\end{equation}
We apply a saddle-point approximation, recognizing that the integral is dominated by times around the minimum of $S_{l,r}(d(\mathbf{r}),\tau)$ at a characteristic time $\tau^*$, such that we approximate $S_{l,r}(d(\mathbf{r}),\tau) = S_{l,r}(d(\mathbf{r}),\tau^*) 
    + \frac{1}{2}(\tau-\tau^*)^2 S_{l,r}''(\tau^*)$, 
with
\begin{equation}
    \tau^* = \left(\frac{C_{l,r}}{d_w-1}\right)^
    {(d_w-1)/d_w}
    d(\mathbf{r})\,\Delta^{-(d_w-1)/d_w}.
\end{equation}
The detailed derivation, including the calculation of $\tau^*$ and the parameter range of validity, is provided in Appendix~\ref{app:saddle}. The value of the action at the saddle point scales as
\begin{equation}
    S(\tau^*) = d_w\!\left(\frac{C}{d_w-1}\right)^{(d_w-1)/d_w}
    d(\mathbf{r})\,\Delta^{1/d_w} \propto \frac{d(\mathbf{r})}{\xi_C},
\end{equation}
and we identify a $C$-dependent localization length
\begin{equation}\label{eq:xi_C}
    \xi_C = \left[\frac{1}{d_w}
    \left(\frac{C_{l,r}}{d_w-1}\right)^{(d_w-1)/d_w}\right]^{-1}
    \Delta^{-1/d_w}.
\end{equation}
Large values of $d(\mathbf{r})$ imply $S_{l,r}(\tau^*)\gg 1$, so the peak is sharp and the approximation is reliable. For small $d(\mathbf{r})$, the peak is broad and the method breaks down. Evaluating the Gaussian integral around $\tau^*$ yields, up to a detuning-dependent prefactor,
\begin{equation}\label{eq:main_result_bs}
    \left| \psi_{\mathrm{BS}}(d(\mathbf{r}))\right| \sim 
    \frac{\exp\!\left[-d(\mathbf{r})/\xi_C\right]}{d(\mathbf{r})^{(d_s-1)/2}}.
\end{equation}
Although eigenstates on fractal lattices are known to exhibit superlocalized, stretched-exponential spatial profiles~\cite{levy1987superlocalization}, the atom-photon bound state seeded by an off-resonant impurity recovers an exponential decay. Since the sub-Gaussian bounds hold with $C_l \leq C \leq C_r$, the localization length is bracketed,
$\xi_{C_r} \leq \xi_C \leq \xi_{C_l}$,
with both bounds sharing the same scaling:
\begin{equation}\label{eq:xi_fractal}
    \xi_{C_{l,r}} \propto \Delta^{-1/d_w}.
\end{equation}
Equation~\eqref{eq:xi_fractal} demonstrates that the scaling of the localization length is controlled by the walk dimension, independently of the precise value of $C$. For regular lattices with $d_w=2$, we recover the well-known scaling $\xi\sim\Delta^{-1/2}$ derived from the effective mass approximation.

We stress that Eq.~\eqref{eq:main_result_bs} is a \textit{pointwise} prediction for a specific emitter--site pair $(\mathbf x_0, \mathbf x)$ at graph distance $d(\mathbf{r}) = d(\mathbf x- \mathbf x_0)$. The actual value of $C$, and therefore the prefactor in $\xi_C$, is a global  constant, but the wavefunction amplitude $\psi(\mathbf{r})$ depends on the full Euclidean position $\mathbf{r}$ of site $x$, not only on $d(\mathbf{r})$. In the bulk of the fractal, many sites share the same chemical distance  $d(\mathbf{r})=r$ from the emitter but occupy very different Euclidean positions, and therefore carry very different wavefunction amplitudes. Averaging $|\psi(x)|^2$ over all such sites at fixed $r$ mixes these different amplitudes, introducing spread into the averaged profile and scatter into the extracted $\xi_C$. Along the outer boundary of the fractal, by contrast, each chemical distance $r$ corresponds to exactly one site and one Euclidean position, so the pointwise prediction is tested directly with no averaging ambiguity. This is why the far-field analysis in Section~\ref{sec:numerics_far_field} is presented for an emitter coupled at the edge of the lattice, at which the exponential decay is cleanest and the extracted localization length is most accurate. The bulk far-field is consistent with the same $\Delta^{-1/d_w}$ scaling but noisier for the geometric reason described above.

\subsection{Near-field scaling}\label{sec:near_field_theory}

The saddle-point approximation used to derive Eq.~\eqref{eq:main_result_bs} assumes that the dominant contribution to the resolvent arises from diffusion times $\tau^\ast \gg 1$, selected by the  factor $e^{-\Delta\tau}$. This approximation is no longer valid in the vicinity of the emitter, where the relevant chemical distances $d(\mathbf{r})$ are small and the associated diffusion times $\tau\sim d(\mathbf{r})^{d_w}$ satisfy $\Delta\tau\ll1$. In this near-field regime, the spatial structure of the bound state is controlled by the short-time behavior of the heat kernel rather than by the exponential decay imposed by the finite detuning. For $d_s<2$ and any fixed detuning $\Delta>0$, the on-site Green's function is finite,
\begin{equation}
|G(0)|\le c_r \int_0^\infty d\tau\,\tau^{-d_s/2}e^{-\Delta\tau}<\infty,
\end{equation}
and diverges only as $\Delta\to0$, consistent with recurrence. The derivation below therefore excludes the  two-dimensional case, for which one recovers a logarithm instead of a power law~\cite{akkermans2012spatial,condamin2007first, grabner1997functional}. The relevant observable is the Green's function difference
\begin{equation}\label{eq:deltaG_def}
\begin{aligned}
    \delta G(d(\mathbf{r})) &\equiv G(d(\mathbf{r}))-G(0)
\\
&= \int_0^{\infty} d\tau\left[K(0,\tau)-K(d(\mathbf{r}),\tau)\right]e^{-\Delta\tau},
\end{aligned}
\end{equation}
which is a standard quantity in diffusion and first-passage-time theory~\cite{condamin2007first,akkermans2012spatial, benichou2008zero, benichou2010geometry}. The detailed derivation, given in Appendix~\ref{app:nearfield}, shows that in the near-field limit $d(\mathbf{r})^{d_w}\Delta\ll1$,
\begin{equation}
\delta G(d(\mathbf{r}))\sim d(\mathbf{r})^{\beta}, \qquad \beta=d_w-d_f,
\label{eq:deltaG_result}
\end{equation}
for $d(\mathbf{r})\gg1$, for which the Alexander--Orbach relation $d_s=2d_f/d_w$~\cite{alexander1982density} has been used and is exact for the nested finitely ramified fractals considered here. Since the bound-state wavefunction is proportional to the Green's function, the near-field amplitude difference obeys
\begin{equation}
|\psi(x_0)-\psi(x)| \propto \delta G(d(x,x_0))\sim d(x,x_0)^{\beta},
\label{eq:near_field_scaling}
\end{equation}
for $d(x,x_0)\ll\xi$, establishing a direct connection to the resistance scaling $R(d(\mathbf{r}))\sim d(\mathbf{r})^{d_w-d_f}$ familiar from classical transport on fractals~\cite{havlin1987diffusion, BarlowBass1990}. This result is interesting in that the photon very close to the emitter obeys a law which incorporates the walk $d_w$ and the fractal dimension $d_f$, both properties which are only defined as global constants and not observable from the propagation over just a few sites. Eq.~\eqref{eq:near_field_scaling} holds in the sense of two-sided bounds: discrete self-similarity leads to log-periodic modulations~\cite{akkermans2012spatial},
$c_1\,d(\mathbf{r})^{\beta}\leq\delta G(d(\mathbf{r}))\leq c_2\,d(\mathbf{r})^{\beta},$
with geometry-dependent constants $c_{1,2}$. On a discrete self-similar fractal, $\delta G(d(\mathbf{r}))$ therefore oscillates log-periodically around the power law $d(\mathbf{r})^\beta$ with an amplitude that is the same for all emitter positions but a phase that depends on where the emitter sits in the fractal's self-similar structure. Averaging over many emitter positions,  these oscillations cancel, revealing the underlying slope $\beta$ on a log-log plot. For $\beta\to0$, we recover the marginal logarithmic scaling of a regular two-dimensional lattice.

\begin{table}[h!]
    \centering
    \renewcommand{\arraystretch}{1.0}
    \setlength{\tabcolsep}{8pt}
    \begin{tabular}{l c c c c}
        \hline
        \textbf{System} & $d_f$ & $d_w$ & $d_s$ & $\beta$ \\
        \hline\hline
        \textbf{1D Chain} & $1.00$ & $2.00$ & $1.00$ & $\mathbf{1.00}$ \\
        \hline
        \textbf{Square Lattice} & $2.00$ & $2.00$ & $2.00$ & $\mathbf{/}$ \\
        \hline
        \textbf{Carpet ($9,1$)} & $1.89$ & $2.10$ & $\approx 1.80$ & $\mathbf{0.20}$ \\
        \textit{\footnotesize{(Infinitely Ramified)}} & & & & \\
        \hline
        \textbf{Carpet ($16,4$)} & $1.79$ & $2.16$ & $\approx 1.66$ & $\mathbf{0.37}$ \\
        \textit{\footnotesize{(Infinitely Ramified)}} & & & & \\
        \hline
        \textbf{Gasket ($b=2$)} & $1.58$ & $2.32$ & $1.36$ & $\mathbf{0.74}$ \\
        \textit{\footnotesize{(Finitely Ramified)}} & & & & \\
        \hline
        \textbf{Gasket ($b=3$)} & $1.63$ & $2.32$ & $1.40$ & $\mathbf{0.69}$ \\
        \textit{\footnotesize{(Finitely Ramified)}} & & & & \\
        \hline
        \textbf{Pyramid ($b=2$)} & $2.00$ & $2.58$ & $1.55$ & $\mathbf{0.58}$ \\
        \textit{\footnotesize{(Finitely ramified)}} & & & & \\
        \hline
        \textbf{Vicsek Fractal} & $1.46$ & $2.46$ & $1.19$ & $\mathbf{1.00}$ \\
        \textit{\footnotesize{(Tree-like)}} & & & & \\
        \hline\hline
    \end{tabular}
    \caption{\label{tab:dimensions} Theoretical dimensional parameters and near-field scaling exponent $\beta=d_w-d_f$. The entries for the Sierpi\'nski carpets and Vicsek fractal are taken from \cite{patino2023brief}, while the gasket and pyramid values follow \cite{hilfer1984renormalisation}. We do not assign $\beta$ to the regular two-dimensional lattice because $d_s=2$ is the marginal logarithmic case.}
\end{table}

\section{Numerical results}\label{sec:five}
In this section we present numerical simulations of atom-photon bound states on self-similar fractal lattices, validating the far-field prediction $\xi\sim\Delta^{-1/d_w}$ and the near-field scaling $\delta\psi\sim d(x,x_0)^\beta$. All results are obtained by exact diagonalization of the coupled emitter-bath Hamiltonian on finite-generation fractal graphs.

\subsection{Far-field localization length}
\label{sec:numerics_far_field}

\begin{figure}
    \centering
    \includegraphics[width=\columnwidth]{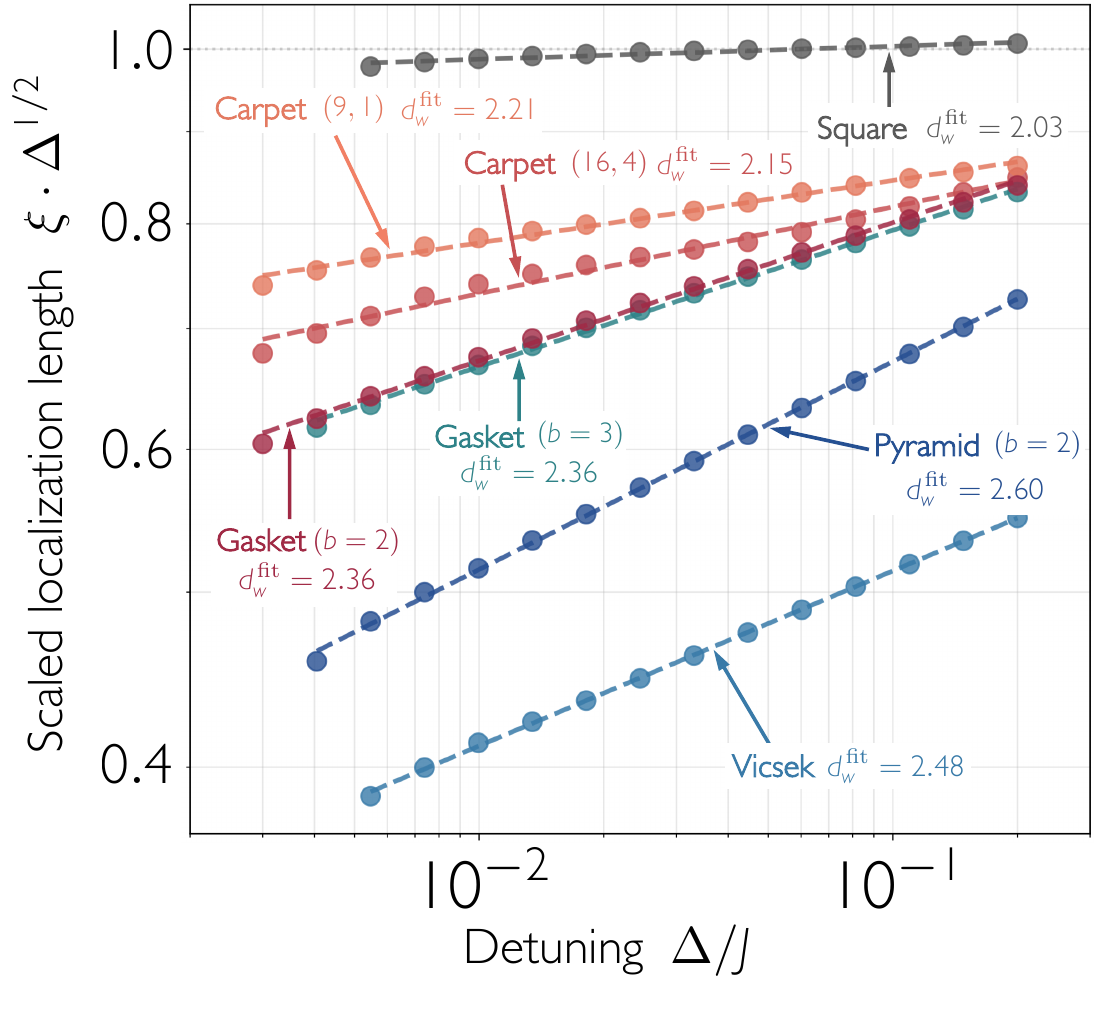}
    \caption{\textit{Far-field localization length.} The rescaled localization length $\xi\cdot\Delta^{1/2}$ is shown as a function of detuning $\Delta$. A horizontal line corresponds to the regular-lattice benchmark $\xi\sim\Delta^{-1/2}$, while systematic power-law drifts reveal the fractal exponent $1/d_w$. Data are extracted along the outer boundary of each fractal, where $d(x,x_0)$ coincides with the Euclidean arc length. The following generations are taken for the fractals: Gasket ($b=2$): $g=7$, Gasket ($b=3$): $g=5$, Carpet ($9,1$): $g=5$, Carpet ($16,4$): $g=4$, Pyramid: $g=6$, Vicsek: $g=5$, Square: $L=151\times 151$ sites.}
    \label{fig:far_field_scaling}
\end{figure}

We extract the localization length $\xi_C$ along the outer boundary of the fractal (parametrized with the coordinate $x$) with the emitter sitting at a site $x_0$. There, the chemical distance $d(x_0,x)$ also coincides with the Euclidean distance as $r=|x-x_0|$, so that the pointwise prediction Eq.~\eqref{eq:main_result_bs} is tested without averaging ambiguity, as argued in Sec.~\ref{sec:four}. To extract $\xi_C$ from the mixed algebraic-exponential form of Eq.~\eqref{eq:main_result_bs}, we perform a linear regression on
\begin{equation}
    \log|\psi_\mathrm{BS}(r)|^2 = \log C - (d-1)\log r - \frac{2}{\xi_C}\,r,
    \label{eq:ols_fit}
\end{equation}
so that the coefficient of the linear term directly yields $\xi_C$. Since the fit quality and the extracted parameters depend sensitively on the chosen window $[r_\mathrm{min}, r_\mathrm{max}]$, we determine the optimal window through a convergence sweep as follows: The upper boundary $r_\mathrm{max}$ is swept from $r_\mathrm{min} + \delta$ to the end of the path, and the regression is performed at each step. The far-field window is identified as the interval over which the fitted exponent $d$ exhibits minimum variance across consecutive windows, indicating that the fit has entered the asymptotic regime in which  Eq.~\eqref{eq:ols_fit} holds. The localization length $\xi_C$ is then taken as the mean over this plateau. We note that $d$ is not further used.

The results are shown in Fig.~\ref{fig:far_field_scaling}. The square lattice yields $d_w^\mathrm{fit} = 2.03$, appearing as a nearly horizontal line and confirming that the rescaling $\xi \mapsto  \xi \cdot \Delta^{1/2}$ correctly removes the $\Delta^{-1/2}$ dependence of the regular-lattice benchmark. Every fractal curve shows a clear upward slope in the log-log plot, confirming $\xi \sim \Delta^{-1/d_w}$ with $d_w > 2$. The fitted walk dimensions agree well with the theoretical values of Table~\ref{tab:dimensions}. We note that we did not consider even smaller detunings due to finite size effects.

\subsection{Near-field amplitude differences}

\begin{figure*}
    \centering
    \includegraphics[width=\textwidth]{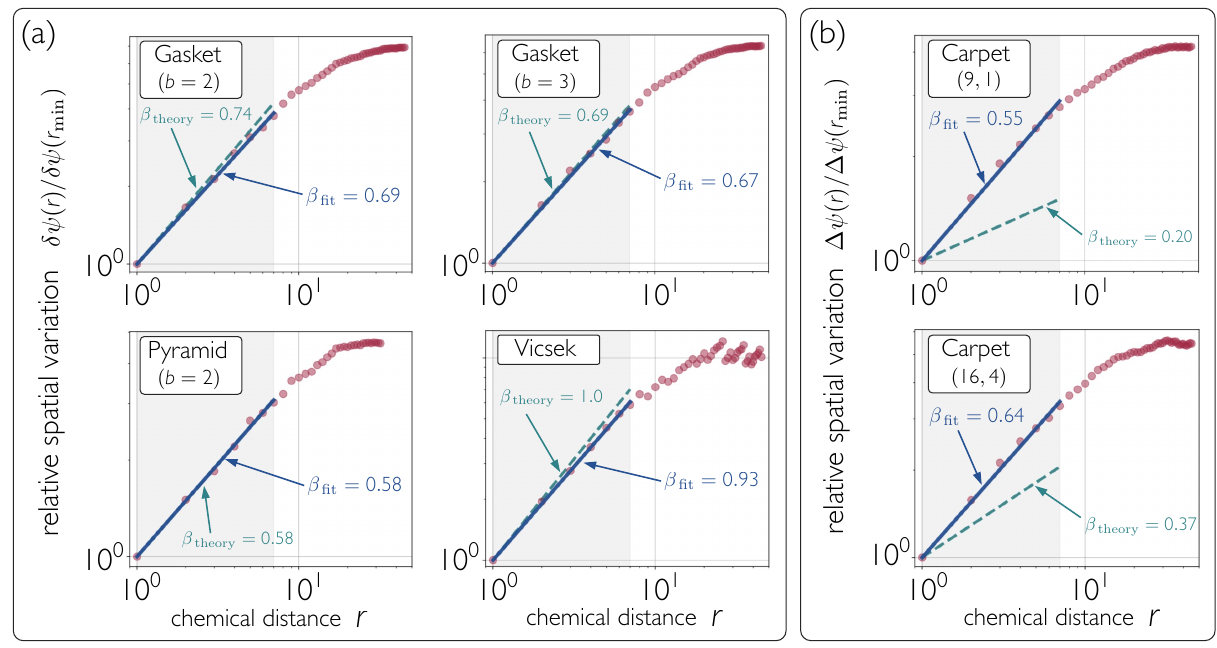}
    \caption{\textit{Near-field scaling of atom-photon bound states}. The quantity $\delta\psi(r)/\delta\psi(r_{\text{min}})$ with $\delta\psi(r) =\big\langle |\psi(\mathbf x_0)-\psi(\mathbf x)| \big\rangle_{r=d(\mathbf x,\mathbf x_0)}$ [red] is shown as a function of the chemical distance $r$ from the emitter together with the fitted exponent $\beta$ [blue] and the theoretical benchmarks [green] from Table~\ref{tab:dimensions}. The gray shaded region highlights the data used for the fit, which is $r\in [1,10]$ for all cases. The nested finitely ramified fractals (a) follow the predicted algebraic law, whereas Sierpi\'nski carpets (b) deviate from it and are numerically closer to the marginal logarithmic behavior of a regular two-dimensional lattice.}
    \label{fig:near_field_scaling}
\end{figure*}

The near-field observable is the averaged amplitude difference $\delta\psi( \mathbf r)=\langle|\psi(\mathbf x_0)-\psi(\mathbf x)|\rangle_{d(\mathbf x,\mathbf x_0)= \mathbf r}$, which corresponds directly to $\delta G(d(\mathbf{r}))$ derived in Sec.~\ref{sec:near_field_theory} and is expected to scale as $d(x,x_0)^\beta$ with $\beta=d_w-d_f$. As established in Eq.~\eqref{eq:deltaG_result}, this scaling holds in the sense of two-sided bounds with geometry-dependent constants $c_1, c_2$. We average over many emitter positions $x_0$ in the bulk to reveal the underlying slope $\beta$. To minimize finite-size effects, only sites whose chemical distance to the outer boundary exceeds a threshold $R_\mathrm{bulk}$ are used for the emitter positions as well as for the positions where the bound state is evaluated. The results are shown in Fig.~\ref{fig:near_field_scaling}. For the nested fractals (Sierpi\'nski gaskets, pyramid, and Vicsek), the data clearly follow the predicted algebraic near-field scaling. For the Sierpi\'nski carpets, the observed behavior deviates systematically from the nested-fractal resistance-scaling benchmark and appears closer to a marginal two-dimensional trend, highlighting the role of infinite ramification and the limited applicability of the Alexander--Orbach-based prediction in this case.

\section{Conclusions}\label{sec:six}
We have shown that atom-photon bound states in self-similar photonic lattices are controlled, in the far field, by the anomalous diffusion exponent of the bath rather than by band curvature. Rewriting the resolvent as a Laplace transform of the heat kernel yields a simple and robust prediction, $\xi\sim\Delta^{-1/d_w}$, which reduces to the familiar regular-lattice law when $d_w=2$. The main physical message is therefore that, in fractal baths, the walk dimension replaces the effective mass as the quantity that sets the range of the dressed light-matter state.

The numerical analysis supports this conclusion across all lattice families considered, provided that the bath Hamiltonian is made equal to the underlying graph Laplacian. This condition can be achieved by adding on-site energies to each site proportional to the number of connections of the site. In contrast, the near-field structure is more geometry dependent. For nested finitely ramified fractals, the amplitude difference follows the predicted algebraic scaling compatible with classical resistance and first-passage arguments. For Sierpi\'nski carpets, however, the short-distance behavior deviates systematically from that simple law and lies closer to the marginal two-dimensional case, highlighting the role of infinite ramification.

Overall, the work identifies fractal waveguide QED as a distinct structured-reservoir regime in which transport exponents leave direct signatures on bound states. Natural extensions include two-emitter interactions, disorder on top of self-similar geometry, and experimental implementations in photonic lattices with engineered onsite potentials. Since fractal reservoirs are generally difficult to treat analytically, our restriction to the lowest spectral edge, implemented through a Laplacian-like lattice, provides a first, relatively clean step toward a systematic understanding of this regime.

\section*{Acknowledgments}
We thank Alejandro Gonz\'alez-Tudela for helpful comments on the manuscript. F.~B.~and F.~K.~K.~acknowledge funding from the Max Planck Society's Lise Meitner Excellence Program 2.0. F.~R.~acknowledges financial support by the European Union-Next Generation EU with the project ``Quantum Optics in Many-Body photonic Environments'' (QOMBE) code SOE2024\textunderscore0000084 -- CUP
B77G24000480006.

\clearpage
\appendix

\begin{widetext}
\section{Saddle-point approximation for far-field scaling}\label{app:saddle}
Starting from Eq.~\eqref{eq:int_1}, we write
\begin{equation}
    -c_r\int_0^\infty d\tau\, \tau^{-d_s/2}\, e^{-S_r(d(\mathbf{r}),\tau)} 
   \le  I(d(\mathbf{r}),\Delta)
   \le -c_l\int_0^\infty d\tau\, \tau^{-d_s/2}\, e^{-S_l(d(\mathbf{r}),\tau)},
\end{equation}
with
\begin{equation}
 S_{l,r}(d(\mathbf{r}),\tau) = \Delta\tau + C_{l,r}\,\left(\frac{d(\mathbf{r})^{d_w}}{\tau}\right)^{1/(d_w-1)}.
\end{equation}
The saddle point $\tau^*$ is obtained from
\begin{equation}
S_{l,r}'(d(\mathbf{r}),\tau)=\Delta-\frac{C}{d_w-1}\left(\frac{d(\mathbf{r})}{\tau}\right)^{\frac{d_w}{d_w-1}}=0 \quad \implies \quad \tau^*=\left(\frac{C_{l,r}}{d_w-1}\right)^{\frac{d_w-1}{d_w}} d(\mathbf{r}) \Delta^{-\frac{d_w-1}{d_w}}.
\end{equation}
At the saddle,
\begin{equation}
    S_{l,r}(d(\mathbf{r}),\tau^*) = d_w\left(\frac{C_{l,r}}{d_w-1}\right)^{\frac{d_w-1}{d_w}} d(\mathbf{r})\Delta^{1/d_w} \propto d(\mathbf{r})\Delta^{1/d_w},
\end{equation}
and
\begin{equation}
S_{l,r}''(d(\mathbf{r}),\tau)=C_{l,r}\,\frac{d_w}{(d_w-1)^2}d(\mathbf{r})^{\frac{d_w}{d_w-1}}\tau^{-\frac{2d_w-1}{d_w-1}}\quad \implies \quad S''(d(\mathbf{r}),\tau^*) = \frac{d_w}{d(\mathbf{r})} \left( \frac{1}{d_w-1} \right)^{\frac{1}{d_w}} \left( \frac{1}{C} \right)^{\frac{d_w-1}{d_w}} \Delta^{\frac{2d_w-1}{d_w}}
\end{equation}
Expanding $S_{l,r}$ to quadratic order around $\tau^*$ and performing the Gaussian integral yields
\begin{equation}
\psi(r)\propto g(\tau^*) \sqrt{\frac{2\pi}{S_{l,r}''(d(\mathbf{r}),\tau^*)}}\, e^{-S_{l,r}(d(\mathbf{r}),\tau^*)}.
\end{equation}
Since $g(\tau^*)\propto d(\mathbf{r})^{-d_s/2}$ and $\sqrt{2\pi/S_{l,r}''(d(\mathbf{r}),\tau^*)}\propto d(\mathbf{r})^{1/2}$ up to detuning-dependent constants, the radial dependence becomes
\begin{equation}
\psi(r)\sim d(\mathbf{r})^{-(d_s-1)/2}\exp\!\left[-\tilde{C}\, d(\mathbf{r})\Delta^{1/d_w}\right],
\end{equation}
which is Eq.~\eqref{eq:main_result_bs} in the main text. The saddle-point approximation requires $S(d(\mathbf{r}),\tau^*)\gg 1$, i.e. $d(\mathbf{r})\gg \xi$, so it is appropriate only for the far-field regime.

\section{Near-field scaling}\label{app:nearfield}
We evaluate the Green's function difference Eq.~\eqref{eq:deltaG_def} in the near-field limit $d(\mathbf{r})^{d_w}\Delta\ll1$. The integrand $K(0,\tau)-K(d(\mathbf{r}),\tau)$ is controlled by the diffusion time required to reach the chemical distance $d(\mathbf{r})$, namely $\tau\sim d(\mathbf{r})^{d_w}$. For $\tau\ll d(\mathbf{r})^{d_w}$ a walker has not yet reached $d(\mathbf{r})$ and $K(d(\mathbf{r}),\tau)\to0$, so the difference reduces to the on-diagonal kernel $K(0,\tau)\sim\tau^{-d_s/2}$. For $\tau\gg d(\mathbf{r})^{d_w}$, the difference is suppressed since $K(d(\mathbf{r}),\tau)\approx K(0,\tau)$. The integral is therefore dominated by $\tau\lesssim d(\mathbf{r})^{d_w}$, for which $e^{-\Delta\tau}\simeq1$ may be dropped.
 
The bounds~\eqref{eq:sub_gaussian_kernel_bounds} depend on position and time only through the combination $d(\mathbf{r})^{d_w}/\tau$. We therefore describe the kernel by the corresponding scaling-form envelope
\begin{equation}\label{eq:scaling_form_app}
    K(d(\mathbf{r}),\tau)=\tau^{-d_s/2}\,
    \Phi\!\left(\frac{d(\mathbf{r})^{d_w}}{\tau}\right),
\end{equation}
with $\Phi$ bounded by~\eqref{eq:sub_gaussian_kernel_bounds} between $c_l\,\exp({-C_l(\cdot)^{1/(d_w-1)}})$ and $c_r\,\exp({-C_r(\cdot)^{1/(d_w-1)}})$, so that $\Phi(0)~>~0$ and $\Phi\to0$ for large arguments. On a discretely self-similar lattice this envelope is approached only up to log-periodic modulations~\cite{akkermans2012spatial}. By substituting $u=\tau/d(\mathbf{r})^{d_w}$ in Eq.~\eqref{eq:deltaG_def}, we factor the distance out of the integral and get
\begin{equation}\label{eq:deltaG_substituted}
    \delta G(d(\mathbf{r}))=I\,d(\mathbf{r})^{\,d_w(1-d_s/2)},
    \qquad \mathrm{with}\quad 
    I=\int_0^\infty du\,u^{-d_s/2}\left[\Phi(0)-\Phi(1/u)\right].
\end{equation}
The integral $I$ is finite. At $u\to0$ the integrand is $\sim u^{-d_s/2}$ and converges because $d_s<2$, while at $u\to\infty$ one has $\Phi(0)-\Phi(1/u)\sim u^{-1/(d_w-1)}$, so that the integrand decays as $u^{-d_s/2-1/(d_w-1)}$ and converges because $d_s/2+1/(d_w-1)>1$ for all fractals in Table~\ref{tab:dimensions}. Its specific value depends on the envelope constants $c_\alpha,C_\alpha$. We can write
\begin{equation}
     \delta G(d(\mathbf{r}))\sim d(\mathbf{r})^{\,d_w(1-d_s/2)}.
\end{equation}
Using the Alexander--Orbach relation $d_s=2d_f/d_w$~\cite{alexander1982density}, the exponent reduces to
\begin{equation}
    d_w\!\left(1-\frac{d_s}{2}\right)=d_w-d_f=\beta,
\end{equation}
so that $\delta G(d(\mathbf{r}))\sim d(\mathbf{r})^{\beta}$, which is the expression~\eqref{eq:deltaG_result}.

\end{widetext}

\bibliographystyle{apsrev4-1}	
\bibliography{references}

\end{document}